\documentclass{emulateapj}
\shorttitle{\sc submillimeter water maser emission in cepheus-a}
\shortauthors{\sc patel et al.}

\begin{document}


\title{Submillimeter Array Observations of 321 GHz Water Maser Emission in Cepheus A}


\author{Nimesh A. Patel\altaffilmark{1}, Salvador Curiel\altaffilmark{2},
Qizhou Zhang\altaffilmark{1}, T. K. Sridharan\altaffilmark{1}, Paul
T. P. Ho\altaffilmark{1,3}, Jos\'e M. Torrelles\altaffilmark{4}}

\altaffiltext{1}{Harvard-Smithsonian Center for Astrophysics,  60 Garden Street, Cambridge, MA; npatel@cfa.harvard.edu, qzhang@cfa.harvard.edu, tksridha@cfa.harvard.edu, pho@cfa.harvard.edu.}
\altaffiltext{2}{Instituto de Astronom{\'\i}a, Universidad Nacional  
Aut\'onoma de M\'exico
(UNAM), Mexico; scuriel@astroscu.unam.mx.}
\altaffiltext{3}{Academia Sinica Institute of Astronomy and Astrophysics, Taipei, Taiwan}
\altaffiltext{4}{Instituto de Ciencias del Espacio (CSIC)-IEEC, Facultat de F\'{\i}sica, 
Universitat de Barcelona, Barcelona, Spain; torrelle@eruopa.ieec.fcr.es}

\begin{abstract}
Using the Submillimeter Array (SMA) we have imaged for the first
time the 321.226 GHz, $10_{29}-9_{36}$ ortho-H$_{2}$O maser emission.
This is also the first detection of this line in the Cepheus A
high-mass star-forming region.  The  22.235 GHz, $6_{16}-5_{23}$
water masers were also observed with the Very Large Array 
43 days following the SMA observations.  Three 
of the nine detected submillimeter maser spots  are associated with
the centimeter masers spatially as well as kinematically, while
there are 36 22 GHz maser spots without corresponding submillimeter
masers. In the HW2 source, both the 321 GHz  and 22 GHz masers occur
within the region of $\sim1$'' which includes the disk-jet system,
but the position angles of the roughly linear structures traced by
the masers indicate that the 321 GHz masers are along the
jet while the 22 GHz masers are perpendicular to it.  
We interpret the submillimeter masers in Cepheus A to be
tracing significantly hotter regions (600$\sim$2000 K) than the
centimeter masers.  \end{abstract}


\section{Introduction}

Since its discovery in the interstellar medium nearly four decades ago
(Cheung et al.  1969), the water maser emission at 22.235 GHz from the
$6_{16}-5_{23}$ transition  has been extensively studied in a variety of
sources, from circumstellar envelopes of late-type stars, star-forming
regions and external galaxies (e.g., Elitzur 1992). This  maser emission
is understood to be arising from shocked gas with temperatures $\sim$500
K and densities  $\sim 10^{9}$ cm$^{-3}$ (Elitzur et al. 1989).

Single-dish observations of the  22 GHz water maser emission have helped
identify the  early stages of massive star-formation (e.g., Sridharan
et al. 2002; Brand et al. 2003). Very Large Array (VLA) and Very Long
Baseline Interferometry (VLBI)  observations have allowed probing of
kinematics of gas in such regions with very high angular resolution of
$0\rlap.{''}08$--0.5 mas (e.g., Claussen et al. 1998; Furuya et al. 2000,
2005; Patel et al. 2000; Imai et al. 2000; Torrelles et al. 2001a,
2001b; Gallimore et al. 2003; Goddi et al. 2005; Goddi \& Moscadelli
2006; Moscadelli et al. 2005).  However, with observations of just
the 22 GHz transition alone, it remains difficult to characterize the
relative differences in the physical conditions in a given star-forming
region. Nearly simultaneous observations of multiple masing transitions at
comparable angular resolutions would be very helpful to allow probing of
physical conditions of the masing gas such as its temperature and density,
beyond just the lower limits required to satisfy the maser excitation.

Except for the 22 GHz transition, and the 183.310 GHz $3_{13}-2_{20}$
transition, other transitions of H$_{2}$O lie in the submillimeter band
(Neufeld \& Melnick 1991). Of these, several submillimeter masers have
been detected with single-dish observations from star-forming regions
as well as late-type circumstellar envelopes (Menten et al.  1990a,
1990b; Melnick et al. 1993). Interferometric observations of some
of these transitions are only now possible with the availability
of the Submillimeter Array (SMA; Ho et al. 2004, Humphreys et
al. 2005).\footnote{ The Submillimeter Array is a joint project between
the Smithsonian Astrophysical Observatory and the Academia Sinica
Institute of Astronomy and Astrophysics, and is funded by the Smithsonian
Institution and the Academia Sinica.}

Cepheus A is a well-studied high-mass star-forming region in the Cepheus
OB3 complex of molecular clouds (Sargent 1977).  Several compact objects
in this region have been identified from radio continuum observations at
centimeter  wavelengths with the VLA (hereafter described as HW sources:
Hughes \& Wouterlout 1984; Garay et al. 1996;  Mart\'{\i}n-Pintado et
al. 2005). The 22 GHz water masers have been observed towards several
of these HW sources (Torrelles et al. 1996, 2001a, 2001b; Vlemmings et
al. 2006) but observations of submillimeter masers have the potential
of allowing a study of differences in the physical conditions  in
these sources, and hence may lead to a better understanding of their
evolutionary differences.

Here we report the first detection of the submillimeter wavelength water
maser emission in the Cepheus-A region. The fact that this emission is
detected at all, suggests the presence of highly excited gas due to the
very high energy levels of the submillimeter maser transitions.

\section{Observations}\label{obs}

The Cepheus A star-forming region was observed with
the SMA  on 2004 August 30 using seven antennas in the
extended configuration with a maximum baseline length of 220
m. The phase center was $\alpha(2000)=22^{h}56^{m}17.971^{s},
\delta(2000)=+62^{\circ}01'49.''279$.  We used a tuning of 321.226
GHz to center the water maser lines in the lower sideband (LSB). The
quasars BL Lac and 0102+584 were observed for 5 minutes each between
every cycle of 20 minutes on the main source.  The spectral band-pass
was calibrated using observations of Uranus and Saturn and absolute
flux calibration was done using Neptune and the quasar 2232+117.
Results from the upper sideband data (centered at 331 GHz, including
CH$_{3}$CN lines and continuum emission) have been reported earlier
by Patel et al. (2005).  Weather conditions were excellent during
the observations with relative humidity $\sim$ 10\% and $\tau_{225
GHz}\approx 0.08$ ($\tau_{321 GHz}\approx 0.69$), measured at the
nearby Caltech Submillimeter Observatory. The track was $\sim$10 hr long
with on-source integration time of $\sim$6.5 hr. $T_{sys,DSB}$ varied
\begin{figure}
\epsscale{.80}
\includegraphics[angle=-90,width=3.5in]{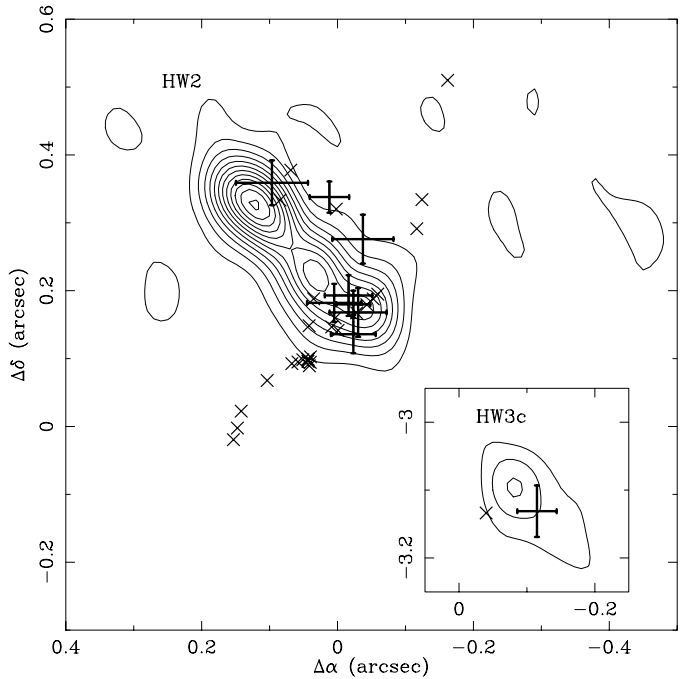}
\caption{VLA map of 1.3 cm continuum emission from HW2 and HW3c
({\it inset}). The positions of submillimeter water masers observed
with the SMA are shown with error bars representing the formal
uncertainties. The position offsets were obtained by fitting
two-dimensional elliptical Gaussians to integrated intensity
maps over channels in which the maser emission appeared (using
the IMFIT task in Miriad). The position offsets are with respect
to the absolute coordinates: $\alpha(2000)=22^{h}56^{m}17.971^{s},
\delta(2000)=+62^{\circ}01'49.''279$. The crosses show the positions of
the 22 GHz water masers observed with the VLA (this Letter).  The errors
in the positions of the 22 GHz water masers are smaller than the size
of the symbols shown.  Only three of these centimeter wavelength masers
appear to be correlated with the submillimeter masers in both position
and velocity (see Table 1).  Only HW2 and HW3c show the presence of
submillimeter water masers among all the compact continuum sources in
the entire Cepheus-A star-forming region (over $\pm 5''$). The VLA
beam is $0.''09\times0.''07$ (P.A. 14$^{\circ}$). The SMA beam is
$0.''8\times0.''7$ with a P.A. of -80$^{\circ}$. }
\end{figure}
from 320 to 500 K. The visibility data were calibrated using the Owens
Valley Radio Observatory's MIR package and imaging was done using the
Miriad package. We applied the self-calibration solutions obtained from
the continuum emission (of $\sim$ 2 Jy beam$^{-1}$) from the HW2 source
obtained from the LSB data, to the spectral-line data of the submillimeter
wavelength water masers.  With uniform weighting, the synthesized beam
was $0.''8\times0.''7$ with a position angle (P. A.) of -80$^{\circ}$.
We expect the maximum error in our absolute astrometry to be $\sim 0.''1$,
based on positions of quasars mapped in our SMA observations. We estimate
an uncertainty of $\sim$20\% in the absolute flux scale in the SMA data.

We also observed the 22.235 GHz water masers in Cepheus-A with the VLA of
the NRAO\footnote{ The National Radio Astronomy Observatory is a facility
of the National Science Foundation operated under cooperative agreement by
Associated Universities, Inc.} in the A configuration on 2004 October 12.
These observations were made in snapshot mode with a total integration
time of about 15 minutes on-source.  The rms noise level in our VLA
observations is about $3.3\times 10^{-3}$ Jy beam$^{-1}$ in channels free
of  maser emission and  $\sim 4.5\times 10^{-1}$ Jy beam$^{-1}$ in the
channel with the strongest maser spot.   The line and continuum data at
1.3 cm were obtained simultaneously and the astrometric registration
between the maser and continuum emission is better than $0.''01$.
These data were reduced using the NRAO AIPS package.
\begin{figure}
\includegraphics[width=3.2in]{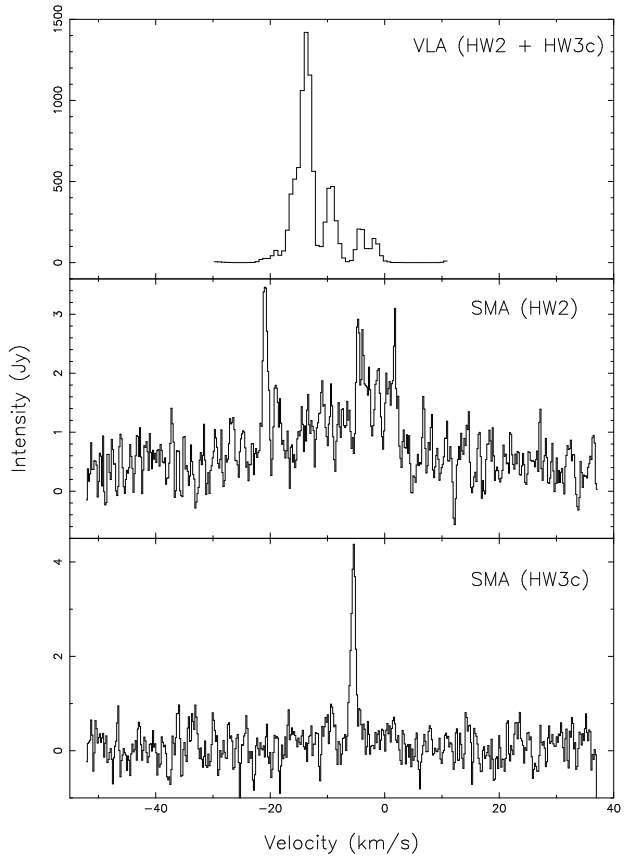}
\caption{Spectra of submillimeter and centimeter wavelength water maser
emission from Cepheus A HW2 and HW3c.}
\end{figure}

\section{Results}\label{results}

Maps of the 1.3 cm continuum emission observed with the VLA (on 2004
October 12), with an overlay of the 22 GHz water masers and the 321 GHz
water masers observed with the SMA are shown in Figure 1. The 321 GHz
water masers were detected only in HW2 and HW3c.  The HW2 thermal jet
has been observed earlier in several epochs from 1991 to 2004 and proper
motions of $\sim$ 500 km s$^{-1}$ of this jet were recently s reported
by Curiel et al. (2006).  The position angle of the jet of $\sim$
45$^{\circ}$ is nearly perpendicular to the proposed circumstellar
disk seen as a flattened elongated structure in continuum emission at
900 $\mu$m and CH$_{3}$CN line emission (Patel et al. 2005). A large
majority of the 22 GHz masers are roughly along the direction of the
disk (which is a larger structure of 1$''$ beyond the scale of Fig. 1),
consistent with that found earlier by Torrelles et al. (1996, 2001a,2001b)
in their VLA and Very Long Baseline Array (VLBA) observations. The 321
GHz masers in HW2 appear to lie along the outflow jet. A total of 39
masers were detected at 22 GHz, only those associated with HW2 and HW3c
are shown in Fig. 1.

Figure 2 shows the spectra of 22 and 321 GHz masers in HW2 and HW3c.
With respect to the systemic velocity of the ambient molecular cloud
in Cepheus A (V$_{LSR}$ $\simeq$  -- 11 km s$^{-1}$; G\'omez et al.
1999), the strongest emission in submillimeter masers associated with
HW2  appears to occur at relatively higher velocities, whereas the 22 GHz
masers are significantly weaker at these velocities.  The brightest 22
GHz masers have flux density of $\sim$1400 Jy, while the brightest 321 GHz
maser is  $\sim$4 Jy. The 22 GHz masers are about a factor of 1000 times
brighter than the 321 GHz masers in general.  The observed characteristics
of the submillimeter masers are listed in Table 1 along with corresponding
values for the three spatially and kinematically associated 22 GHz masers.

\section{Discussion}\label{discussion}

From the sizes of the 321 GHz emission spots in channel maps and the
observed flux densities, we estimate a lower limit in the brightness
temperature to be $\sim$200 K. Since we do not have a reliable estimate
of the sizes of these spots, we are unable to check whether the emission
is due to maser action based on the brightness temperature. The narrow
line-widths of the 321 GHz emission (0.3--1 km s$^{-1}$) compared to
the widths of the CH$_{3}$CN lines, strongly suggest that the 321 GHz
lines are indeed due to maser emission.

The first astronomical detection of the $10_{29}-9_{36}$ transition
of H$_{2}$O masers was made by Menten et al. (1990a) with the Caltech
Submillimeter Observatory 10.4 m telescope.  Shortly after the discovery
of the 321 GHz masers, Neufeld \& Melnick (1990, hereafter, NM90), showed
that the observed 22  and 321 GHz masers can be explained by collisional
excitation within the same volume of gas. NM90 presented models of the
emissivity ratio R of 22 GHz / 321 GHz maser luminosities as a function
of $\xi$ and T, where $\xi$ is a combination of parameters involving the
gas density, abundance of water and the magnitude of velocity gradient in
the slab of gas that is masing (see equation 1 of NM90).  In a subsequent
paper, Neufeld \& Melnick (1991) presented more detailed theoretical
calculations including 349 rotational states of ortho and para water under
a range of $\xi$ and radiation field.  The 321 GHz transition is among the
20 other theoretically predicted maser transitions in the frequency range
of 183--1542 GHz. Many of these transitions are shown to be approaching in
brightness to the 22 GHz maser, under certain ranges of T and $\xi$. Yates
et al. (1997) carried out similar calculations for several submillimeter
H$_{2}$O maser transitions, including the 321 GHz line.  Their range of
physical conditions is wider than in NM90 and Neufeld \& Melnick (1991),
and they also considered the effects of radiation from dust grains.
According to these models, the 321 GHz maser is strongly inverted under
more restricted conditions than the 22 GHz line (see Fig. 3 of NM90 and
Fig. 4a of Yates et al. 1997). In fact, strong inversion of the 321 GHz
transition requires  T$_{k}>1000$K. Assuming a temperature of 1400 K,
the greatest gain in the 321 GHz line occurs for gas density n(H$_{2}$)
in the range of 4--6$\times$10$^{8}$cm$^{-3}$ (Yates et al. 1997).

All of these model calculations have a central assumption that the
submillimeter and centimeter wavelength  masing transitions arise in the
same volume of gas.  This implies that the 22 GHz masers have a broader
range of physical conditions in temperature and densities (Yates et al.
1997).  (There can be 22 GHz emission and no accompanying 321 GHz masers
in the same region but not vice-versa, according to these models). We can
confront these models with our maps of 22 and 321 GHz masers, to first
ask if these masers arise in the same volume of gas, and if they do,
what are the implications of the observed luminosity ratio.

Figure 1 and Table 1 show that we have two out of eight masers in HW2
that appear to be coincident in both spatial and velocity neighborhood
($<0.''1$ and $<3$ km s$^{-1}$) with 22 GHz masers.  In addition, we have
one more region in the HW3c which is located $\sim$3$''$ south of HW2,
that appears to have masers at both 22 GHz and 321 GHz arising from
the same region. If we use a more stringent criteria for coincidence
of masers in line-of-sight velocity to be within 1 km s$^{-1}$, based
on the typical line widths of the maser emission, we would have only
one maser spot (occurring in HW3c; number 9 in Table 1) that would be
coincident. The coincidence of masers 1 and 7 in HW2, might therefore
be considered to be  tentative, given the larger difference in their
velocities. We speculate that this difference in velocity is related to
the fact that HW2 is a dynamically more active region compared to HW3c.

The emissivity ratio $R$ (22/321 GHz) for these masers listed as numbers
1, 7, and 9 in Table 1 are  9.2, 0.02, and 0.18.  The value of $R$ is
roughly expected to decrease with temperature, given that $E/k=643$K
for the 22 GHz transition and 1861 K for the 321 GHz line. In this way,
a much hotter gas is required for the submillimeter wavelength  maser
excitation.  Assuming that both 22 and 321 GHz masers are saturated, the
luminosity ratio is  then the same as the ratio of rate coefficients as
defined by NM90 and plotted in their Fig. 3.  We find that for maser-1
(Table 1), the observed ratio implies the kinetic temperature to be in
the range of 500--2000K, depending on the value of $\xi$. According to
NM90, the typical value of log $\xi$  would be in the range of -1.25
to -0.25 for the shocked gas in star-forming regions and their Figure 3
would then imply a temperature of $\sim$ 600K for the region where maser 1
occurs. Masers 7 and 9 have much smaller values of $R$, much less than 1.
Even for $R=5$, according to NM90, such masers would presumably exist
behind slower non-dissociative shocks in which the kinetic temperature
of the gas is $\sim$ 1000K.  Again, according to figure 3 of NM90, this
value of R $<$ 1 implies gas temperatures as high as 2000K for a wide
range of $\xi$.

With respect to the remaining six masers (listed as numbers 2--6 and
8 in Table 1) that show emission only in the 321 GHz and not in the 22
GHz transition, we cannot apply any of the existing theoretical models
to these $R \rightarrow 0$ cases that imply that the 22 GHz masers are
quenched but the 321 GHz emission remains strong. According to Figure
3 of NM90, which has the lowest contour corresponding to $R=1.5$, the
regions with $R \rightarrow 0$ may imply temperatures greater than 2000K.
On the other hand, we cannot completely rule out  the effect of time
variability for the absence of 22 GHz masers corresponding to some the
observed 321 GHz masers.

The 22 GHz masers are well known to be time-variable.  Previous VLBA
observations of Cepheus A (Torrelles et al. 2001a, 2001b) and IRAS
21391+5802 (Patel et al. 2000), show that several masing spots disappear
over timescales of 1 month. These masers however, tend to be relatively
weaker. The theoretical models as in NM90 predict strong emission at
22 GHz corresponding to the 321 GHz masers. If these corresponding 22
GHz masers existed during the epoch of the submillimeter observations,
they are unlikely to have been completely extinguished about a  month
later in the centimeter wavelength observations.

Much less is known about the variability of the 321 GHz masers.
In late-type stars, near simultaneous observations of 22, 321, and 325
GHz masers were carried out by Yates \& Cohen (1996) over four epochs
separated by $\sim$ 20 days. They find the 321 GHz masers to be the
most variable,  in terms of both changes in brightness -- by a factor
of 2 -- 10 and the shortness of time scale, by $\sim$20 days.  However,
it remains unclear how far we can carry forward these findings to the
case of water masers associated with star-forming regions.

The 22 GHz masers that are not associated with the 321 GHz masers are
likely to be arising in relatively cooler regions  among the various
HW sources in Cepheus A, compared to HW2 and HW3c.  The fact that the
submillimeter wavelength masers in HW2 are found along the major axis of
the jet, lead us to propose that they are arising in hotter gas regions
due to the impact of the jet.  

\section{Conclusions}\label{conclusions}

We report the first detection of 321 GHz water masers in the high-mass
star-forming region Cepheus A. The SMA observations at $\sim0.''8$ have
sufficient spatial resolution to allow a study of the association of these
masers with the multiple radio continuum sources such as HW2 and HW3c
(separated by about $3''$), and also with the disk-outflow system of HW2
which has a size scale of about $1''$ (725 AU). We had near-simultaneous
22 GHz water maser observations made with the VLA that show, in general,
a poor agreement between the locations of submillimeter and centimeter
water masers: only three out of nine masers agree in both position and
line-of-sight velocities.  Moreover, the submillimeter masers are found
to lie along the outflow jet whereas the 22 GHz masers are primarily
found perpendicular to it.  The mapping of submillimeter wavelength
masers with the SMA, along with near-simultaneous mapping of the 22
GHz water masers with the VLA, opens the possibility of studying the
physical conditions of density and temperature of the gas associated
with high-mass star-forming sites (where conditions are likely to be
suitable for excitation of submillimeter water masers), at very high
spatial resolution ($\sim$ 100~AU).

\acknowledgments
It is a pleasure to thank James Moran, Gary Melnick,  David Neufeld and
Elizabeth Humphreys  for helpful discussions. We are grateful to the SMA
staff in Cambridge, Hilo and Taipei, for their help with observations.
We thank the referee for helpful comments on the manuscript.
SC acknowledges support from DEGAPA/UNAM and from CONACyT (M\';exico)
grant  43120$-$F. JMT acknowledges partial financial support from the
Spanish grant AYA2005-08523-C03.

\begin{deluxetable}{ccccccccc}
\tablecolumns{9}
\tablewidth{0pt}
\tablecaption{Summary of detected masers\label{tbl-1}}

\tablehead{
\multicolumn{1}{c}{}&
\multicolumn{4}{c}{321 GHz {\sc masers}\tablenotemark{a}}&
\multicolumn{4}{c}{22 GHz {\sc masers}\tablenotemark{b}}\\
\multicolumn{1}{c}{}&
\multicolumn{4}{c}{\hrulefill}&
\multicolumn{4}{c}{\hrulefill}\\
\multicolumn{1}{c}{No.} &
\multicolumn{1}{c}{$\Delta\alpha$} &
\multicolumn{1}{c}{$\Delta\delta$} &
\multicolumn{1}{c}{Flux} &
\multicolumn{1}{c}{} &
\multicolumn{1}{c}{$\Delta\alpha$} &
\multicolumn{1}{c}{$\Delta\delta$} &
\multicolumn{1}{c}{Flux} &
\multicolumn{1}{c}{} \\
\multicolumn{1}{c}{} &
\multicolumn{1}{c}{} &
\multicolumn{1}{c}{} &
\multicolumn{1}{c}{Density} &
\multicolumn{1}{c}{$V_{LSR}$} &
\multicolumn{1}{c}{}&
\multicolumn{1}{c}{} &
\multicolumn{1}{c}{Density} &
\multicolumn{1}{c}{$V_{LSR}$} \\
\multicolumn{1}{c}{} &
\multicolumn{1}{c}{$''$} &
\multicolumn{1}{c}{$''$} &
\multicolumn{1}{c}{(Jy)} &
\multicolumn{1}{c}{(km s$^{-1}$)}&
\multicolumn{1}{c}{$''$} &
\multicolumn{1}{c}{$''$} &
\multicolumn{1}{c}{(Jy)} &
\multicolumn{1}{c}{(km s$^{-1}$)}
}
\startdata
1 & 0.096 &  0.359  & 2.8 & -19.0 &0.068&0.378&25.9&-21.1\\
2 & 0.012 &  0.338  & 4.8 & -21.0&...&...&...&...\\
3 & -0.037 & 0.276   & 2.9 & -11.0&...&...&...&...\\
4 & -0.016 &  0.193  & 2.7 &  -4.5 &...&...&...&...\\
5 & 0.005 &  0.182  & 3.0 & -1.0 &...&...&...&...\\
6 & -0.023 &  0.180  & 6.5 & -3.5 &...&...&...&...\\
7 & -0.030 &  0.168  & 5.1 & 1.2&-0.004&0.167&0.09&4.0\\
8 & -0.023 &  0.136  & 1.7 &  -3.8 &...&...&...&...\\
9\tablenotemark{c} & -0.115 &  -3.131  & 3.7 &  -5.5 &-0.005&-3.133&0.67&-5.6\\
\enddata

\tablenotetext{a}{Typical uncertainty in SMA flux values is 0.4 Jy. Formal
uncertainty in Gaussian fitted positions is $\sim$ 30 mas (see error-bars
on position offsets showned  in Fig. 1).}

\tablenotetext{b}{The 22 GHz masers listed here are a subset of all the
masers detected in the VLA observations. Only those masers which are most
likely to be associated with the 321 GHz masers (closest occurrences in
position and velocity) are tabulated here. Typical uncertainty in VLA flux
values is 10 mJy and formal uncertainty in positions is $\sim$ 1 mas.}

\tablenotetext{c}{This maser is associated with HW3c, all others are
associated with HW2.}

\end{deluxetable}

\end{document}